\documentclass[%
 reprint,
 amsmath,amssymb,
pra,
floatfix,
]{revtex4-1}
\usepackage{amsmath}
\usepackage{amssymb}
\usepackage{graphicx}
\usepackage{placeins}
\usepackage{caption}
\usepackage{subcaption}
\usepackage{hyperref}
\usepackage{braket}
\usepackage{cleveref}
\usepackage{mathrsfs}
\usepackage{graphicx}
\usepackage{dcolumn}
\usepackage{bm}
\usepackage{hyperref}
\usepackage[mathlines]{lineno}
\usepackage{color}
\begin{document}
\preprint{APS/123-QED}
\title{Polarizability Extraction for Waveguide-Fed Metasurfaces}

\author{Laura Pulido-Mancera}
 \email{laura.pulido@duke.edu}
\affiliation{%
Department of Electrical and Computer Engineering, Duke University, Durham, NC, 27708.\\
Kymeta Corporation, 12277 134th Court NE, Redmond, Washington 98052, USA
}
\thanks{This work was supported by the Air Force Office of Scientific Research (AFOSR, Grant No. FA9550-12-1-0491)}
\author{Patrick T. Bowen}
\affiliation{%
 Department of Electrical and Computer Engineering, Duke University, Durham, NC, 27708.\\
 Kymeta Corporation, 12277 134th Court NE, Redmond, Washington 98052, USA
}
\author{Mohammadreza F. Imani}
\affiliation{%
 Department of Electrical and Computer Engineering, Duke University, Durham, NC, 27708.
}
\author{Nathan Kundtz}
\affiliation{%
Kymeta Corporation, 12277 134th Court NE, Redmond, Washington 98052, USA
}
\author{David Smith}
\affiliation{%
Department of Electrical and Computer Engineering, Duke University, Durham, NC, 27708.
}

\begin{abstract}
We consider the design and modeling of metasurfaces that couple energy from guided waves to propagating wavefronts. This is a first step towards a comprehensive, multiscale modeling platform for metasurface antennas---large arrays of metamaterial elements embedded in a waveguide structure that radiates intro free-space---in which the detailed electromagnetic responses of metamaterial elements are replaced by polarizable dipoles. We present two methods to extract the effective polarizability of a metamaterial element embedded in a one- or two-dimensional waveguide. The first method invokes surface equivalence principles, averaging over the effective surface currents and charges within an element to obtain the effective dipole moments; the second method is based on computing the coefficients of the scattered waves within the waveguide, from which the effective polarizability can be inferred. We demonstrate these methods on several variants of waveguide-fed metasurface elements, finding excellent agreement between the two, as well as with analytical expressions derived for irises with simpler geometries. Extending the polarizability extraction technique to higher order multipoles, we confirm the validity of the dipole approximation for common metamaterial elements.  With the effective polarizabilities of the metamaterial elements accurately determined, the radiated fields generated by a metasurface antenna (inside and outside the antenna) can be found self-consistently by including the interactions between polarizable dipoles. The dipole description provides an alternative language and computational framework for engineering metasurface antennas, holograms, lenses, beam-forming arrays, and other electrically large, waveguide-fed metasurface structures.
\end{abstract}

\pacs{Valid PACS appear here}
\maketitle


\section{\label{sec:Introduction} Introduction}
The concept of metamaterials---artificially structured materials---has prompted the rapid development of new tools and techniques for controlling the propagation of waves. The metamaterial paradigm, in which an artificial medium is assembled from a collection of subwavelength elements with desired scattering characteristics, has had a profound impact across numerous scientific fields, including physics \cite{tretyakov2003analytical,simovski2007magnetic}, electromagnetic \cite{engheta2006metamaterials} and acoustic \cite{popa2009design} wave phenomena, materials science, chemistry, engineering \cite{caloz2005MMtransmissionline}, and nanoscience \cite{fedotov2006asymmetric,kildishev2013planar}. Metamaterials have provided a venue to tailor material properties in ways not feasible with conventional materials \cite{smith2004metamaterials,pendry2006controlling,capolino2009theory}, opening the door to unique and often exotic wave phenomena such as negative and near-zero refractive index materials  \cite{smith2000negative,eleftheriades2005negative,padilla2006negative}. Furthermore, metamaterials research has lead to the demonstration of unprecedented devices, such as transformation optical structures and invisibility cloaks \cite{schurig2006metamaterial}, as well as superlenses\cite{fang2003imaging}.  

The underlying metamaterial paradigm is that the behavior of waves propagating within a large (many wavelengths) composite medium can be understood from the properties of constituent elements---each subwavelength in dimensions---and their mutual interactions. The advantage of the metamaterial perspective is that the properties of each of the constituent elements can be determined exactly using a full-wave simulation over a relatively small, subwavelength, spatial domain. From these simulations, effective constitutive parameters can be retrieved, replacing the detailed current and field distributions within the small domain by just a few parameters such as the electric permittivity and magnetic permeability \cite{chen2004robust}. The wave propagation properties of the composite structure can then be modeled by solving Maxwell's equations directly, with effective constitutive parameters replacing the actual metamaterial structures. While the effective constitutive parameters obtained by numerical retrieval methods must be applied with considerable caution, retrieval methods have nevertheless been used with success in the design of many metamaterial structures  \cite{smith2002retrieval,smith2005retrieval}. Replacing the details of an artificial medium with effective constitutive parameters facilitates device simulations and optimization cycles, vastly reducing the computational requirements since the individual elements are replaced by homogenized constitutive parameters. Complex metamaterial devices have been designed and demonstrated by this technique, including the transformation optical structures that rely on precise variations in material properties throughout a volume \cite{luneburg2010,ni2012broadband,aieta2015multiwavelength}.

Despite the compelling features of volumetric metamaterials and their unique and often unprecedented material properties, the applications for volumetric metamaterials have been limited. This limitation exists because some of the most intriguing properties of metamaterials occur near resonances in the individual elements---resonances that often impose bandwidth limitations and produce large resistive losses. Thus, waves propagating through any significant volume (even just a few wavelengths) of a metamaterial can be heavily attenuated. In addition, fabricating metamaterial elements to control electric and magnetic fields polarized in arbitrary directions, and assembling such elements throughout a volume, remains a major implementation challenge; typically, the properties of volumetric metamaterials have been demonstrated in highly constrained formats and proof-of-concept prototypes.

The difficulties associated with volumetric metamaterials are considerably reduced for metamaterial structures consisting of just a single layer or a few layers of metamaterial elements---also known as metasurfaces \cite{hollowaysmith2012metasurface}. Being easier to design, model and implement \cite{holloway2011characterizing,holloway2012use,yu2014flat}, metasurfaces have rapidly gained traction as a major subfield in metamaterials research \cite{glybovski2016metasurfaces}. As quasi-optical devices, metasurfaces provide control of reflection and transmission across the spectrum \cite{pors2013broadband}, paving the way for advanced components such as flat lenses\cite{yu2014flat,yu2011light}, thin polarizers\cite{zhao2011manipulating,baena2015self}, spatial or frequency filters \cite{xue2010new}, and holographic and diffractive elements \cite{aieta2015multiwavelength,dupre2014grating,aieta2012aberration}. Used as coatings, metasurfaces can control the absorbance and emissivity of a surface, and thus have relevance to thermophotovoltaics \cite{bauer2011thermophotovoltaics}, detectors and sources \cite{landy2008perfect,albooyeh2012huge,teperik2007total,chen2006active,ra2015thin,gu2010broadband,grady2013terahertz}. Given the capabilities of metasurfaces to control waves, but without many of the limitations of volumetric metamaterials, metasurfaces have proven a good match for commercialization efforts, with many serious applications now being pursued, including satellite communications \cite{johnson2014discrete,stevenson2016mtenna, sleasman2017experimental}, radar \cite{pulido2016rma}, and microwave imaging \cite{lipworth2015virtualizer,hunt2013science,hunt2014metaimager}.

As with a volumetric metamaterial, the scattering properties of a metasurface can be characterized by a set of effective surface constitutive properties, which homogenize---or average over---the properties of many identical, discrete, metamaterial elements \cite{epstein2016huygens,kuester2003averaged}. These effective medium properties relate to the discontinuity of the fields across the metasurface---approximated as having infinitesimal thickness---and are encapsulated in a set of generalized boundary conditions \cite{pfeiffer2013metamaterial,wong2014design, yu2011light}. Effective surface constitutive properties are a convenient and natural description for free-standing metasurfaces illuminated by plane waves \cite{pfeiffer2013cascaded,zhao2011homogenization,sievenpiper2002tunable,sievenpiper2003two}; however, there are many contexts where a homogenization description is not the most convenient or the most accurate. For example, a metasurface interacts with a guided wave in ways not easily captured with a simple homogenized description. To better illustrate this point, consider a rectangular waveguide with one bounding surface patterned with a collection of metamaterial elements, each consisting of a pattern of voids or irises in the conducting surface. The volume of the waveguide just beneath the element behaves as if filled with a material possessing effective constitutive parameters introduced by the presence of the metamaterial element \cite{landy2013homogenization}. In addition, the waveguide-fed metamaterial elements also leak energy out of the waveguide, such that the entire structure can act as an aperture antenna \cite{hunt2014metaimager,mario1767metasurfing}. Indeed, metasurfaces can provide nearly total control over the wavefront in a manner similar to phased arrays \cite{hansen2009phasedarray} and other aperture antennas, but often with advantages not available in other formats \cite{sievenpiper2005forward,minatti2012circularly,patel2013modeling,johnson2015sidelobe,quarfoth2013artificial}. 

For metasurfaces, a weakness associated with homogenization techniques is that the element size and average spacing between metamaterial elements must be significantly subwavelength---a condition not necessarily satisfied in many situations. An element size of one-tenth to one-fifth of a wavelength is typical for many metamaterial structures, which implies the phase of the wave will have significant variation over the volume containing the element. Such metamaterials are said to exhibit spatial dispersion, and are properly characterized by constitutive parameters that depend on the wave vector in addition to the frequency, adding complication to the homogenization description. While numerical retrievals include spatial dispersion in the effective constitutive parameters, the retrieved parameters are specifically valid only in the exact arrangement simulated. For example, if the effective constitutive parameters are retrieved from a simulation of a cubic cell with periodic boundary conditions, the retrieved parameters will only be specifically valid for that medium, and not necessarily applicable when the same element is placed in a different context---for example, in a random arrangement.

To arrive at a more generally valid description of a metamaterial while still avoiding a full-wave simulation of the composite structure, we consider directly the properties of each metamaterial scattering element. The response of such a metamaterial element can generally be expressed in a series of induced electric and magnetic multipoles, typically dominated by the dipole term. The strength of the dipolar contribution is connected to an effective polarizability, which represent the coupling between the total dipole moment and the incident field on the element. To the extent that the higher order multipoles beyond the dipolar term can be neglected, the scattering from a collection of equivalent dipoles provides a near exact and computationally efficient model of the metamaterial structure. By assigning an effective polarizability to a metamaterial element rather than treating the metamaterial or metasurface as a continuous medium with constitutive parameters, it is possible to predict the overall response of the structure without any limitation on the element's periodicity or arrangement. The combination of polarizability extraction and the dipole representation forms an alternative, powerful modeling platform for metasurfaces and metamaterials. 

To obtain the polarizability of an arbitrary metamaterial element, a technique that can be applied is to first assume the element is part of an infinitely periodic medium \cite{karamanos2012polarizability,scher2009extracting,belov2005homogenization}. The details of the periodic structure can then be replaced by periodic boundary conditions, so that the full-wave simulation domain extends only over a single cell of the structure. An effective polarizability of the element, which includes the contributions from all other elements in the infinite array, can then be extracted from the computed field or charge/current distributions. Finally, the intrinsic polarizability can be determined using the Lorentz formula that relates the intrinsic and effective polarizabilities \cite{belov2005homogenization}. 

Our aim here is to extend the dipole model as an analytical tool for waveguide-fed metasurfaces. Unlike the free-standing metasurface or volumetric metamaterial, for which each metamaterial element can be reduced to a free space dipole, an individual metamaterial element patterned in a waveguide also interacts with the waveguide structure. Even if the metamaterial element is reduced to a single free space dipole, the image dipoles induced within the waveguide structure must then be taken into account to obtain an accurate expression for the waveguide-fed metamaterial element \cite{tretyakov1991averaging,feresidis2005artificial}. Rather than working through this complication, we can instead apply a numerical polarization extraction procedure using the waveguide modes, arriving at an effective polarizability that includes all of the waveguide interactions. Once having established this dipole description of a waveguide-based metamaterial element, we can subsequently find the properties of the composite waveguide-fed metasurface accurately, efficiently and quickly, for example using the dipole model previously described.

We approach this problem as follows: In Section \ref{sec:DDA} we introduce the polarizability framework for a metamaterial element embedded in a waveguide and summarize the self-consistent dipole model for metasurfaces. In Section \ref{sec:DirectIntegration} we apply the equivalence principle to the waveguide-fed metamaterial, and derive integrals relating the equivalent current densities to the effective dipole moments. The results of the direct integration method are used as the basis for comparison with the second method we present in Section \ref{sec:PolarizabilityExtraction}, in which the polarizability is obtained using the scattering (S) parameters of the element when placed in a waveguide. Both methods have been used in the context of numerical retrieval of effective constitutive parameters for volumetric metamaterials, with the former method related to field averaging \cite{smith2000averaging}, while the latter method related to the well-known S-parameters retrieval method \cite{smith2002retrieval,smith2005retrieval}. In Section \ref{sec:SimulatedResults} we perform polarizability extractions for different waveguide-fed metasurface geometries: circular iris \cite{Bethe}, elliptical iris \cite{collin1991field,elliott1975design}, iris-coupled patch antenna \cite{buck1986aperture} and the complementary electric inductive-capacitive metamaterial resonator(cELC) \cite{schurig2006electric}. We extend the polarizability extraction method to two-dimensional (2D-) waveguide structures in section (\ref{sec:Planar-Waveguide}), where the S-parameters cannot be used to retrieve polarizability. To address this problem, we outline an alternative method based on the mode expansion of cylindrical waves propagating through the waveguide. This framework is particularly advantageous for modeling and designing planar structures \cite{hunt2014metaimager,sleasman2015DyAp1}. We also demonstrate that the extracted polarizability of a metamaterial element depends on the geometry of the waveguide in which it is embedded. As a means of confirming the accuracy of the dipole approximation, we express the induced fields from a metamaterial element as a multipole expansion, comparing the relative strengths of the expansion coefficients. The result of this analysis shows that, indeed, the dipole term dominates the response, justifying the dipolar description and use of the dipole model. We conclude by examining the potential application of the polarizability extraction for different metasurface-based devices.

\section{Self-consistent Dipole Model for Metasurfaces \label{sec:DDA}}

Our conceptual picture of a waveguide-fed metasurface is that of a collection of polarizable dipoles, each of which accounts for the scattering associated with an actual metamaterial element. Using the polarization extraction technique to be presented, a metamaterial element is reduced to an effective electric dipole moment and an effective magnetic dipole moment, $\mathbf{p}$ and $\mathbf{m}$, which are proportional to the local electric or magnetic field at the center of the element multiplied by a coupling coefficient, termed the dynamic polarizability \cite{liu2016polarizability}. It is referred to as ``dynamic" since it describes the element's response due to a time-varying incident field. A time-dependent electric field can induce solenoidal currents and thus can give rise to a magnetic polarization in addition to the electric polarization. To properly take into account the co- and cross-coupling of the electromagnetic fields excited within the metamaterial element, the dynamic polarizability should be represented as a tensor, so that $\mathbf{p}= \bar{\bar{\alpha}}_{ee}\mathbf{E}^{loc}$ and  $\mathbf{m} = \bar{\bar{\alpha}}_{mm}\mathbf{H}^{loc}$, where $\mathbf{H}^{loc}$ and $\mathbf{E}^{loc}$ are the local electric and magnetic fields at the center of the element.  

It is worth noting that the polarizability of a metamaterial element is not an inherent property, but rather depends on the environment, and therefore may be considered a nonlocal property. To better illustrate this point, consider a single passive magnetic dipole $\textbf{m}$ placed at position \(\textbf{r}_0\) in an environment that is described by the Green's function \(\textbf{G}(\textbf{r},\textbf{r}')\). This configuration is illuminated by an incident magnetic field \(\textbf{H}^{0}\). The total field \(\textbf{H}^{tot}\) everywhere throughout the environment is thus
\begin{equation}\label{eq:TotalH}
\textbf{H}^{tot}=\textbf{H}^{0}+\textbf{G}(\textbf{r},\textbf{r}_0)\textbf{m}.
\end{equation}

\noindent There are two possible ways to define the polarizability of a dipole \cite{bowen2017effective}. One is the effective polarizability, defined in terms of the incident wave as \(\textbf{m}=\alpha_m\textbf{H}^{0}\), and the other is the inherent polarizability, defined in terms of the total field \(\textbf{m}=\tilde{\alpha}_m\textbf{H}^{tot}\). The latter case would be a difficult definition to uphold consistently for a point particle, because the real part of the Green's function \(\textbf{G}(\textbf{r},\textbf{r}_0)\) diverges as \(\textbf{r}\rightarrow\textbf{r}_0\). (Some authors have developed ways of dealing with the singularity and relate it to the radiation of the dipole \cite{sipe1974macroscopic,sondergaard2004surface}.) 

The imaginary part of the Green's function at the origin, however, is well-known to have an important physical meaning related to the radiation reaction force on the dipole \cite{novotny2012principles,bowen2017effective}. Defining the polarizability as \(\textbf{m}=\alpha_m\textbf{H}^{0}\), we consider the average power absorbed by the dipole according to Poynting's theorem: 
\begin{equation}\label{eq:PoyntingTheoremAbsorption}
P_{abs}=(1/2)\int\mathrm{Re}\left\{\textbf{J}_m^*\cdot\textbf{H}\right\}\mathrm{d}V
\end{equation}
\noindent where \(\textbf{J}_m=i\omega\mu_0\textbf{m}\delta^{(3)}(\textbf{r}-\textbf{r}_0)\) is the magnetic current associated with the magnetic dipole,  $\omega$ is the angular frequency, $^{*}$ represents complex conjugate, and $\mu_0$ is the permeability of free space. If there are no Ohmic losses in the system, then all power scattered by the dipole is re-radiated, and consequently \(P_{abs}=0\). Since the magnetic field \(\textbf{H}\) that appears in Eq. \ref{eq:PoyntingTheoremAbsorption} is the total magnetic field given in Eq. \ref{eq:TotalH}, \(P_{abs}=0\) implies that
\begin{equation}
\mathrm{Im}\left\{\textbf{m}^*\cdot\textbf{H}^{0}-\textbf{m}^*\textbf{G}(\textbf{r},\textbf{r}_0)\textbf{m}\right\}=0.
\end{equation}
Substituting \(\textbf{m}=\alpha_m\textbf{H}^{0}\) and assuming that the polarizability is isotropic, we obtain a relationship between the imaginary part and the real part of the polarizability as
\begin{equation}
\mathrm{Im}\{\alpha_m\}=|\alpha_m|^2\mathrm{Im}\{\textbf{m}^*\textbf{G}(\textbf{r}_0,\textbf{r}_0)\textbf{m}/|\textbf{m}|^2\}, \label{eq:pol-imagpart}
\end{equation}
\noindent which must be satisfied in order for the conservation of energy to be upheld. Equation (\ref{eq:pol-imagpart}) implies that \(\mathrm{Im}\left\{1/\alpha_m\right\}=\mathrm{Im}\{\textbf{m}^*\textbf{G}(\textbf{r}_0,\textbf{r}_0)\textbf{m}/|\textbf{m}|^2\}\). Therefore, after some simple algebraic manipulation, Eq. \ref{eq:pol-imagpart} can be recast as  
\begin{equation}
\alpha_m=\frac{\tilde{\alpha}_m}{1+i\tilde{\alpha}_m\mathrm{Im}\{\textbf{m}^*\textbf{G}(\textbf{r}_0,\textbf{r}_0)\textbf{m}/|\textbf{m}|^2\}}.\label{eq:general_rad_reaction}
\end{equation}
Here, \(\tilde{\alpha}_m\) is the inherent polarizability of the dipole, whereas \(\alpha_m\), which is defined as the ratio between the dipole moment and the incident field, is dependent on the local environment. If a dipole is placed in free space, then a Taylor series expansion of the Green's function shows that \(\mathrm{Im}\{\textbf{G}(\textbf{r}_0,\textbf{r}_0)\}=k^3/6\pi\), and this yields the radiation reaction to the polarizability of a dipole in free space
\begin{equation}\label{eq:FreeSpacePolarizability}
\alpha_m=\frac{\tilde{\alpha}_m}{1+i\tilde{\alpha}_mk^3/6\pi}.
\end{equation}

\noindent The expression in Eq. \ref{eq:FreeSpacePolarizability} is often known in the literature as the radiation reaction correction or the Sipe-Krankendonk relation \cite{sipe1974macroscopic,belov2003condition,strickland2015dynamic}. If instead the dipole is placed just above an infinite ground plane, the dipole radiates twice as much energy, and so \(\mathrm{Im}\{\hat{\textbf{m}}\textbf{G}(\textbf{r}_0,\textbf{r}_0)\hat{\textbf{m}}\}=k^3/3\pi\). If the dipole is a complementary metamaterial element embedded in a waveguide wall, then it will radiate both into the upper half space and into the waveguide, and so the radiation reaction correction would need to take into account both scattered fields. Considering that this correction must account for half of the radiation in free-space, as in Eq. \ref{eq:FreeSpacePolarizability}, and half of the radiation inside the waveguide, as derived in appendix \ref{ap:rad-reaction}, the corrected polarizability has the form

\begin{equation}
\alpha_m=\frac{\tilde{\alpha}_m}{1+i\tilde{\alpha}_m (k^3/3\pi + k/ab)}.\label{eq:waveguide_rad_reaction}
\end{equation}

Equation (\ref{eq:waveguide_rad_reaction}) has significant implications on any polarizability extraction method that deals with waveguide integrated metamaterial elements. For example, if the static polarizability of an element is calculated using Bethe theory (as discussed in section \ref{sec:SimulatedResults}), then the radiation reaction correction will be different depending on whether that element is placed in a 2D waveguide or a cavity, and so the proper correction as in Eq. \ref{eq:general_rad_reaction} will need to be applied in each environment. 

Once the polarizabilities are determined, an accurate, self-consistent description of the scattering from the collection of dipoles can be obtained using the dipole model--including the Green's function associated with the specific environment--which expresses the relationship between the local fields $\mathbf{E}^{loc}(\mathbf{r}_i)$ in terms of the incident fields $\mathbf{E}^{0}(\mathbf{r}_i)$ and the fields scattered from all of the dipoles. These equations can be written as
\begin{equation}
\mathbf{E}^{loc}(\mathbf{r}_i) =\mathbf{E}^{0}(\mathbf{r}_i) + 
\sum_{j\neq i}  \mathbf{G}_{ee}(\mathbf{r}_i,\mathbf{r}_j) \mathbf{p}(\mathbf{r}_j) + \mathbf{G}_{em}(\mathbf{r}_i,\mathbf{r}_j) \mathbf{m}(\mathbf{r}_j)
\label{eq:DDA-eq}
\end{equation}
\noindent where $\mathbf{G}_{ee}, \mathbf{G}_{em}$ correspond to the electric components of the dyadic Green's function. An equivalent expression can be derived for the magnetic field. Examining the expression in Eq. \ref{eq:DDA-eq}, it can be seen that these coupled equations capture the interaction of the incident wave with each of the meta-atom (through the $j=0$ terms) as well as interaction between different elements (the summation term). Note that we do not solve these equations in the present work, but include them to provide the complete modeling framework. The development of the waveguide Green's function and simulation of various waveguide-fed metasurfaces will be presented separately.

\section{Polarizability extraction in a rectangular waveguide: Direct Integration}\label{sec:DirectIntegration}

\begin{figure}
\centering{\includegraphics[width=0.95\columnwidth]{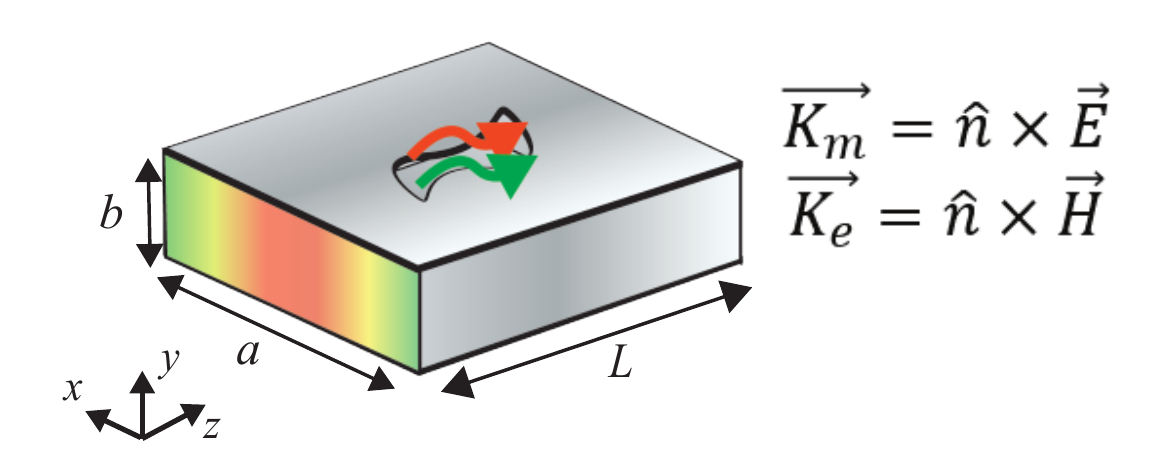}}
\caption{Metamaterial element etched on a rectangular waveguide. The waveguide is excited by the fundamental mode $TE_{10}$ and the metamaterial element induces electric and magnetic currents.}\label{fig-DDA_for_RWG_int}
\end{figure}

We start by considering an arbitrarily shaped iris etched into the upper conducting surface of a rectangular waveguide, as shown in Fig.\ref{fig-DDA_for_RWG_int}. The coordinate system is chosen so that the propagation direction is in the \(z\)-direction; \(a\) corresponds to the width of the waveguide along the \(x\)-axis, and \(b\) corresponds to the height along the \(y\)-axis. The guided wave couples to the iris, which radiates a portion of the incident wave into the free space region. A common methodology to solve the radiated field by such a structure is the surface equivalence principle. This principle states that the electric field on the boundary of a domain can be represented as a magnetic surface current \(\textbf{K}_m=\textbf{E}\times\hat{\textbf{n}}\), while the magnetic field on the boundary can be represented as an electric surface current \(\textbf{K}_e=\hat{\textbf{n}}\times\textbf{H}\), where $\mathbf{E}$ and $\mathbf{H}$ are the total fields on the surface of the domain, and $\hat{\mathbf{n}}$ is the normal to the surface. Using $\textbf{K}_m$ and $\textbf{K}_e$ and the corresponding Green's functions, one can determine the field within the domain. 

Applying this principle to the geometry of Fig. \ref{fig-DDA_for_RWG_int}, we see that the tangential electric field is zero everywhere on the waveguide surface except over the void regions defining the iris; if the iris is deeply subwavelength, then the field scattered into the far-field may be approximated by just the first term of the multipole expansions of \(\textbf{K}_e\) and \(\textbf{K}_m\). Hence, the dipole moments representing the iris can be calculated as \cite{Jackson}
\begin{subequations}
\begin{align}
\mathbf{p} = \epsilon_0 \hat{\textbf{n}} \int \textbf{r} \cdot \mathbf{E}^{tan} da \\
\mathbf{m} = \frac{1}{i \mu \omega} \int \hat{\textbf{n}} \times \mathbf{E}^{tan} da .
\end{align}\label{eq:peff-meff}
\end{subequations}
\noindent The integration is performed over the surface of the iris, $\hat{\textbf{n}}=\hat{\textbf{y}}$ is the vector normal to the top surface, and $\mathbf{E}_{tan}$ corresponds to the tangential field at the surface of the iris. It is worth noting the tangential magnetic field is not zero over the surface of waveguide; however, the tangential magnetic field corresponds to an electric current density parallel to a metallic wall, and by image theory its effect can be ignored. 

Since the effective dipole moments are proportional to the incident fields $\mathbf{E}^0,\mathbf{H}^0$, we can define the effective polarizabilities as
\begin{equation}\label{eq:alpha-effs}
 \mathbf{p} = \epsilon_0  \bar{\bar{\alpha}}^e_{eff} \mathbf{E^0} \quad \mathbf{m} = \bar{\bar{\alpha}}^m_{eff} \mathbf{H^0}. 
 \end{equation}

\noindent In the most general case, each iris can be described by an electric polarizability tensor $\bar{\bar{\alpha}}^e_{eff}$ and a magnetic polarizability tensor $\bar{\bar{\alpha}}^m_{eff}$. These tensors are symmetric and would thus normally have six unknown parameters, each of which would need to be extracted from a full-wave simulation. In free space, these components can be found by computing the scattered fields in all directions. In the waveguide geometry considered here, such simulations are not possible. Instead, we take advantage of the geometry of the iris and the symmetry of the excitation field into account. For example, when the symmetry axes of iris are coincident with the symmetry axes of the waveguide, we can assume that the polarizability tensor is diagonal. Further, the boundary conditions of the waveguide require that the incident tangential electric field \(E_x^0\) and \(E_z^0\) and the normal component of the magnetic field \(H_y^0\) be equal to zero on the surface of the waveguide. Therefore, the tensor components \(\alpha_{px}\), \(\alpha_{pz}\), and \(\alpha_{my}\) can never be excited, and hence we can assume them to be zero. The polarizability tensors then reduce to \(\bar{\bar{\alpha}}^e_{eff}=\mathrm{diag}(0,\alpha_{py},0)\) and \(\bar{\bar{\alpha}}^m_{eff}=\mathrm{diag}(\alpha_{mx},0,\alpha_{mz})\). Hence, the polarizability extraction is simplified to finding three unknowns (\(\alpha_{py}\), \(\alpha_{mx}\), \(\alpha_{mz}\)). One further unknown can be removed if the iris is placed where the magnetic field of the incident mode has a null in the \(z\)-component. In that case, only \(\alpha_{py}\) and \(\alpha_{mx}\) are relevant to the problem. By using Eq. \ref{eq:peff-meff} and Eq. \ref{eq:alpha-effs} and the previously described assumptions, the characteristic polarizabilities are
\begin{subequations}
\begin{align}
\alpha_{ey} = \frac{\int\int(xE_x + zE_z)dxdz}{E^{0}_{y}} \\
\alpha_{mx} = \frac{1}{i \mu \omega H^{0}_x} \int \int E_z dxdz .
\end{align}\label{eq:pol_love}
\end{subequations}
Equations Eq. \ref{eq:pol_love} provide a simple method to calculate the polarizability of an element from a full-wave simulation of the fields of the element embedded in a waveguide structure.

\section{Polarizability extraction in a rectangular waveguide: Scattering Parameters}\label{sec:PolarizabilityExtraction}

While calculating the polarizability of a metamaterial element by means of Eq. \ref{eq:pol_love} provides a physically accurate characterization, the integration over the surface of the element can be cumbersome to perform for all desired frequency points and for arbitrary geometries. In many instances, this integration may also be subject to numerical inaccuracies due to singularities near edges or coarse meshing, as it is especially the case for resonant elements such as those examined in section \ref{sec:SimulatedResults}. Instead of the direct integration, in this section we consider the extraction of the polarizabilities from the fields scattered by the element into the waveguide. For this calculation, we apply coupled mode theory to determine the coupling of the element embedded in a waveguide to the forward and backward scattered fields within the waveguide. The fields inside the waveguide at any plane of constant \(z\) (along the propagation direction) can be expanded as a discrete sum of orthogonal modes. These modes are defined as  \cite{Jackson}
\begin{subequations}\label{eq:total-Efields}
\begin{align}
\textbf{E}_n^+&=\left(\textbf{E}_{n t}(x,y)+\textbf{E}_{n z}(x,y)\right)e^{-i\beta_n z} \\
\textbf{H}_n^+&=\left(\textbf{H}_{n t}(x,y)+\textbf{H}_{n z}(x,y)\right)e^{-i\beta_n z} \\
\textbf{E}_n^-&=\left(\textbf{E}_{n t}(x,y)-\textbf{E}_{n z}(x,y)\right)e^{i\beta_n z} \\
\textbf{H}_n^-&=\left(-\textbf{H}_{n t}(x,y)+\textbf{H}_{n z}(x,y)\right)e^{i\beta_n z}
\end{align}
\end{subequations}
\noindent where \(\textbf{E}^-_n\) and \(\textbf{E}^+_n\) are respectively the waveguide modes traveling in the backwards and forwards directions. The subscript ``\(t\)" refers to the component of the fields that are transverse to the direction of propagation, and $\beta_n$ is the propagation constant of the $n$th mode. The mode normalization used in Eq. \ref{eq:total-Efields} is defined from the integral over the cross section of the waveguide, such that 
\begin{equation}\label{eq:ModeNormalizationE}
\int\textbf{E}_n\cdot\textbf{E}_m\mathrm{d}a=\delta_{mn},
\end{equation}
\noindent where $\delta_{mn}$ is 1 for $n=m$ and 0 otherwise. Note that the electric fields of the modes are dimensionless, and the magnetic fields have units of inverse impedance as
\begin{equation}\label{eq:ModeNormalizationH}
\int\textbf{H}_n\cdot\textbf{H}_m\mathrm{d}a=\delta_{mn}/Z_n^2,
\end{equation}
\noindent where the wave impedance \(Z_n\) is defined as a normalization constant for each  mode as
\begin{equation}\label{eq:PowerNormalization}
Z_n=\frac{1}{\int\textbf{E}_n\times\textbf{H}_n\cdot\hat{\textbf{n}} da}.
\end{equation}

\begin{figure}
\centering{\includegraphics[width=0.85\columnwidth]{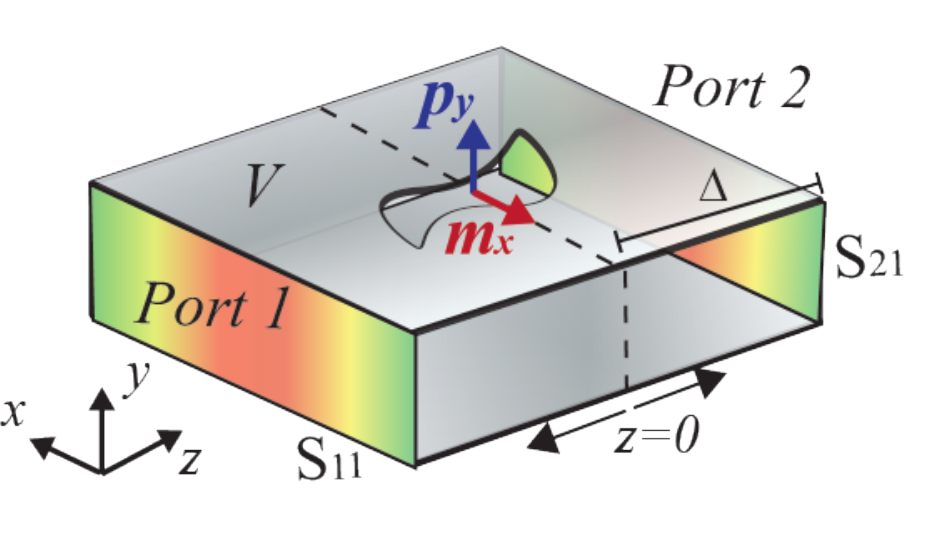}}
\caption{Metamaterial element effectively acts as a electric and magnetic dipole that scatters inside the waveguide.}\label{fig-DDA_for_RWG}
\end{figure}

In Eq. \ref{eq:PowerNormalization} the integration is over the cross sectional surface of the waveguide, i.e. the surface representing \textit{Port 1} in Fig. \ref{fig-DDA_for_RWG}. 

Consider a metamaterial element placed at the center of the top plate of the waveguide. We assume that the incident field is the forward-propagating fundamental mode---coming from \textit{Port 1}---with unit amplitude \(\textbf{E}^{0+}\), as shown in Fig. \ref{fig-DDA_for_RWG}. When the metamaterial element is present, it couples and scatters to all modes. While the element has a finite size, for points inside the waveguide that are few wavelengths away, the element is well-approximated as a point scatterer placed at z=0 (location of the element). In the absence of the metamaterial element, the total field is simply the incident field, which is identical in both the forward and backward directions. As a result, we express the modal decomposition of the total (both incident and scattered) fields into backwards propagating modes at \textit{Port 1} ($z=-\Delta$) as
\begin{equation}\label{eq:Ep}
\textbf{E}^{-}=\textbf{E}_0^++\sum_n A_n^-\textbf{E}_n^-, 
\end{equation}
where \(\textbf{A}^-_n\) are the amplitudes of the modes scattered by the element in the backwards direction, and $n$ is the mode number. Similarly, a modal decomposition of the fields in the plane of \(z=+\Delta\) into forward propagating modes yields
\begin{align} \label{eq:Em}
\textbf{E}|^{+}&=\textbf{E}_0^++\sum_n A_n^+\textbf{E}_n^+
\end{align}
\noindent where \(\textbf{A}^+_n\) are likewise the mode amplitude coefficients of the scattered field by the element in the forward direction, and the incident field \(\textbf{E}_0^+\) has been written as a separate term. In these calculations, $\Delta$ can be any distance as long as it is larger than the size of metamaterial element.

We first consider a volume within the waveguide that encompasses the metamaterial element (bounded by the two ports). The incident field impinging on the element will induce a set of fields that we denote as $\textbf{E}$ and $\textbf{H}$. Within the coupled mode formulation, these fields can be related to the waveguide modes through Poynting's theorem, or 
\begin{equation}\label{eq:poynting}
\nabla \cdot (\mathbf{E} \times \mathbf{H}^{(\pm)}_n - \mathbf{E}^{(\pm)}_n \times \mathbf{H})= \textbf{J}_e\cdot\textbf{E}^{(\pm)}_n-\textbf{J}_m\cdot\textbf{H}^{(\pm)}_n.
\end{equation}
We integrate Eq. \ref{eq:poynting} over the volume $V$ the portion of the waveguide between the two ports, and applying the divergence theorem, Eq. \ref{eq:poynting} becomes
\begin{align}\label{eq:poyntingS}
&\int_S (\mathbf{E} \times \mathbf{H}^{(\pm)}_n - \mathbf{E}^{(\pm)}_n \times \mathbf{H})\cdot \hat{\mathbf{n}} da = \nonumber \\ 
&\int_V \textbf{J}_e\cdot\textbf{E}^\pm_n-\textbf{J}_m\cdot\textbf{H}^\pm_n dV
\end{align}
\noindent where $S$ is the closed surface that encloses $V$ and $\hat{\mathbf{n}}$ is an outwardly directed normal. Since the waveguide walls are assumed to be perfectly conducting, the only nonzero contributions to the surface integrals arise from the surfaces representing \textit{Port 1} and \textit{Port 2} (depicted in Fig.\ref{fig-DDA_for_RWG}), and the surface of the metamaterial element. Since the field $\textbf{E}$ can be written as a fictitious magnetic surface current through $\textbf{K}_m=\textbf{E}\times\hat{\textbf{n}}$ and $\textbf{H}$ can be related to a fictitious electric surface current in the same way, then the surface integral in Eq. \ref{eq:poyntingS} indicates the manner in which the effective dipoles representing the metamaterial element couple to each of the waveguide modes, as might be expected from Lorentz reciprocity. 

To obtain the amplitude coefficients $A_n^{(\pm)}$, we assume there are no current sources in the volume, implying that the volume integral in Eq. \ref{eq:poyntingS} vanishes. Substituting the expansions of the fields in Eq. \ref{eq:Em} and Eq. \ref{eq:Ep} into Eq. \ref{eq:poyntingS} and using the orthogonality relations in Eq. \ref{eq:ModeNormalizationE,eq:ModeNormalizationH}, we obtain the scattered amplitude coefficients as an overlap integral of the waveguide mode fields with the total field taken over the surface of the metamaterial element. More explicitly, the amplitude coefficients can be found as 
\begin{equation}\label{eq:Apm_integral}
A_n^{(\pm)}= \frac{Z_n}{2}\int_{element} (\mathbf{E} \times \mathbf{H}^{(\mp)}_n-\mathbf{E}^\mp_n \times \mathbf{H})\cdot \mathbf{n} da.
\end{equation} 
In an alternative approach, the electric field in the aperture could be considered zero and replaced by an equivalent electric and magnetic surface current, according to the equivalence principle. In this case, the surface integral vanishes everywhere except over the surfaces of the ports, but the volume integral over the meta-atom becomes a surface integral of the equivalent surface currents \(\textbf{K}_m=\textbf{E}\times\hat{\textbf{n}}\) and \(\textbf{K}_e=-\textbf{H}\times\hat{\textbf{n}}\). Using Eq. \ref{eq:Ep} and Eq. \ref{eq:Em} and invoking orthogonality, we obtain
\begin{equation}\label{eq:Apm_integral2}
A_n^{(\pm)}= \frac{Z_n}{2}\int_{element} (\mathbf{K}_e \cdot \mathbf{E}^{(\mp)}_n-\mathbf{K}_m \cdot \mathbf{H}^{(\mp)}_n)da.
\end{equation} 
Since the metamaterial element is deeply subwavelength, the fields of the waveguide modes can be expanded in a Taylor series around the center of the element. The lowest order term is constant over the surface of the element, yielding
\begin{align}\label{eq:Apm_integral3}
A_n^{(\pm)}= \frac{Z_n}{2} \Bigg[&\mathbf{E}^{(\mp)}_n(\textbf{r}_0)\cdot\int_{element} \mathbf{K}_e da\nonumber \\
&-\mathbf{H}^{(\mp)}_n(\textbf{r}_0)\cdot\int_{element}\mathbf{K}_m da\Bigg].
\end{align} 
As previously stated in Eq. \ref{eq:peff-meff}, the two integrals in Eq. \ref{eq:Apm_integral3} are proportional to the electric and magnetic dipole moments, \(\textbf{p}\) and \(\textbf{m}\). Therefore, the final expression for the amplitude coefficients in terms of these dipole moments is given by
\begin{subequations}
\begin{align}
A_n^+=\frac{i\omega Z_n}{2}\left(\textbf{E}^-_n\cdot\textbf{p} - \mu_0 \textbf{H}^-_n \cdot\textbf{m} \right) \label{eq:ForwardScatteringA} \\
A_n^-=\frac{i \omega Z_n}{2}\left(\textbf{E}^+_n\cdot\textbf{p} - \mu_0 \textbf{H}^+_n\cdot\textbf{m}\right). \label{eq:BackwardScatteringA}
\end{align}\label{eq:Ap-Am}
\end{subequations}
Equation \ref{eq:alpha-effs} shows that the dipole moments are related to the incident fields, which in turn can be expanded in terms of eigenmodes. Since the incident field is the fundamental mode, the polarizability is defined by
\begin{subequations}
\begin{align}
\textbf{p}=\epsilon \bar{\bar{\alpha}}^{eff}_e\textbf{E}_0^+ \\
\textbf{m}=\bar{\bar{\alpha}}^{eff}_m\textbf{H}_0^+.
\end{align}
\end{subequations}
Due to the symmetry of the fields in the rectangular waveguide, the \(\alpha_{mz}\) component cannot be excited by the \(z\)-component of the magnetic field. Hence, Eq. \ref{eq:Ap-Am} reduces to two coupled equations with two unknowns: \(\alpha_{py}\) and \(\alpha_{mx}\), which can be recast as
\begin{subequations}
\begin{align}
A_n^+=\frac{i\omega Z_n}{2}\left(\epsilon \alpha_{py}E^+_{0y}E^-_{ny} - \mu_0 \alpha_{mx}H^+_{0x}H^-_{nx}\right) \label{eq:Ap-pols} \\
A_n^-=\frac{i \omega Z_n}{2}\left(\epsilon \alpha_{py}E^+_{0y}E^+_{ny} - \mu_0 \alpha_{mx} H^+_{0x}H^+_{nx}\right) \label{eq:Am-pols}
\end{align}\label{eq:Ap-Am-2}
\end{subequations}
Considering the orthogonality of the eigenmodes and the symmetry properties of the electromagnetic fields---the transverse components of the electric field are symmetric under a flip of direction (i.e. \(E_{ny}^-=E^+_{ny}\)), while the magnetic field is antisymmetric (i.e. \(H_{nx}^-=-H^+_{nx}\))---we can solve Eq. \ref{eq:Ap-Am-2} in order to find the polarizabilities as
\begin{subequations}
\begin{align}
\alpha_{py}= \frac{2}{i\omega Z_n} \frac{(A_0^+ + A_0^-)}{\epsilon (E^+_{0y})^2}\\
\alpha_{mx}= \frac{2}{i\omega Z_n} \frac{(A_0^+ - A_0^-)}{\mu (H^+_{0x})^2}.
\end{align}\label{eq:pol_AmAp}
\end{subequations}
For the fundamental mode, the normalized fields and impedance at the dipole location are given by
\begin{equation} 
 |\textbf{E}^+_{0y}|^2= \frac{4}{ab} \quad |\textbf{H}^+_{0x}|^2= \frac{4\beta_{10}^2}{ab Z_0^2 k^2} \quad Z_{0} = \eta k/\beta_{10} \label{eq:ALLfields}
\end{equation}
\noindent where $\eta$ is the vacuum impedance. Furthermore, the amplitude coefficients $A^+_0$ and $A^-_0$ correspond to the amplitude terms for the fundamental mode of the scattered fields in the forward and backward directions. Therefore they are directly related to the scattering parameters with respect to each port: the reflected field, related to $A^-_0$ is proportional to the reflection coefficient i.e. $S_{11}$, while the transmitted field related to $A^+_0$ is proportional to the transmission coefficient $S_{21}$ and the incident field in the forward direction. More explicitly, these relationships are expressed by
\begin{equation}\label{eq:An-sparam}
A^-_0= S_{11} \quad A^+_0= S_{21}-1.
\end{equation}
Taking into account Eq. \ref{eq:An-sparam} in conjunction with Eq. \ref{eq:pol_AmAp} and Eq. \ref{eq:ALLfields} it is possible to find the final expression for the polarizabilities as 
\begin{subequations}
\begin{align}
\alpha_{py}&=\frac{-iab\beta_{10}}{2k^2}(A_0^++A_0^-) = \frac{-iab\beta_{10}}{2k^2}(S_{21}+S_{11}-1) \\
\alpha_{mx}&=\frac{-iab}{2\beta_{10}}(A_0^+-A_0^-) = \frac{-iab}{2\beta_{10}}(S_{21}-S_{11}-1) .
\end{align}\label{eq:pol_rwg}
\end{subequations}
Equation \ref{eq:pol_rwg} provides the polarizabilities of any metamaterial element embedded in a rectangular waveguide in terms of the scattering parameters, which can be obtained from direct measurement or full-wave simulation. However, it is important to note that the scattering parameters must be de-embeded to the plane of the metamaterial element before using them in Eq. \ref{eq:pol_rwg}. This is a straightforward process well-known in literature \cite{karamanos2012polarizability}. Another important point to note is that we have assumed ports which only excite/represent single mode. This condition should be applied when simulating these structures in numerical solvers. More importantly, since it is cumbersome in experiment to excite purely the fundamental mode, the ports should be placed at least a few wavelengths in distance away from the metamaterial element to ensure the non-propagating higher order modes have decayed. 

The equations in Eq. \ref{eq:pol_rwg} are similar to the expressions found for the effective polarizabilities of metamaterial elements in periodic metasurfaces \cite{karamanos2012polarizability,scher2009extracting}. This relationship means that a single element in a rectangular waveguide acts as a dipole whose response is equivalent to the response of the element in a periodic metasurface. The details of this equivalence will be discussed in a future paper.

\section{Simulated Results for the Rectangular Waveguide}\label{sec:SimulatedResults}

Using a full-wave simulation, we can extract the polarizability of arbitrary metamaterial elements patterned into rectangular waveguides from the two different approaches described in sections \ref{sec:DirectIntegration} and \ref{sec:PolarizabilityExtraction}. For both extraction techniques, a single full-wave simulation in \textit{CST Microwave Studio} is performed assuming a waveguide designed to operate over frequencies in the X-band (8-12 GHz). The waveguide dimensions are $a=21.94 \;\mathrm{mm}$, $b=5 \;\mathrm{mm}$, $L=22.7 \;\mathrm{mm}$ and thickness $1.27 \;\mathrm{mm}$. We perform this simulation for several different metamaterial element geometries: circular iris, elliptical iris, iris-coupled patch antenna, and the cELC resonator. 

In addition to the methods described above, the dipole moments of simple geometries, such as an elliptical iris may be also obtained from the static dipole moments of general ellipsoidal dielectric and permeable magnetic bodies. Consider an elliptically shaped aperture with the major axis along the $x-$direction and minor axis along the $z-$direction. Let the major radius be $l_1$ and minor radius $l_2$. In the static limit, the polarizabilities of such an elliptical iris are given by \cite{collin1991field}
 \begin{subequations}
 \begin{align}
 \tilde{\alpha_{mx}} = \frac{4\pi l_1^3 e^2}{3[E(e) - K(e)]}\\
 \tilde{\alpha_{mz}} = \frac{4\pi l_1^3 e^2(1-e^2)}{3[E(e) - (1-e^2)K(e)]}\\
 \tilde{\alpha_{ey}}= -\frac{4\pi l_1^3 (1-e^2)}{3E(e)},
 \end{align}\label{eq:pol-ellipse}
 \end{subequations}
\noindent where $e= \sqrt{1-(l_2/l_1)^2}$ (assuming $l_1>l_2$) is the eccentricity of the ellipse, and $K(e)$ and $E(e)$ are the complete elliptic integrals of the first and second kind, respectively. If $e=0$ these expressions reduce to the static polarizabilities of circular irises. Using these static expressions for the dynamic polarizability will violate conservation of energy since it lacks the radiation damping term of dynamic polarizability. These expressions may be corrected by the radiation term, as described in Eq. \ref{eq:waveguide_rad_reaction}, assuming the real part given in Eq. \ref{eq:pol-ellipse}. 

\begin{figure}
\centering{
\includegraphics[width=\columnwidth]{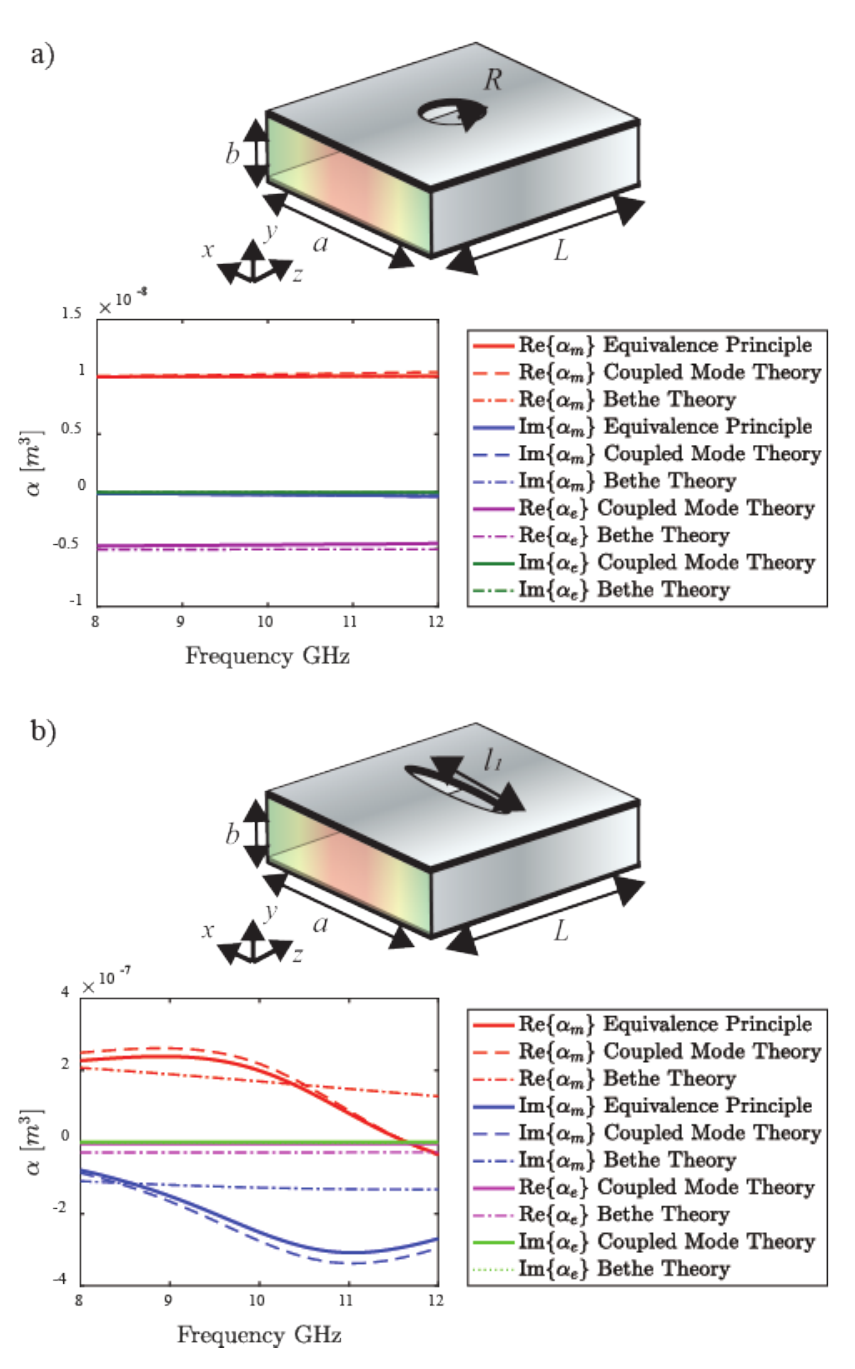}}
\caption{Dynamic Polarizability of small apertures. a) Circular iris. Dimensions are $R=2$ mm. b) Elliptical iris. Dimensions are $l_1=7.5 mm$ and $l_2=0.5 mm$.}\label{fig:pol_results_rwg}
\end{figure}

Figure \ref{fig:pol_results_rwg} shows the polarizability of simple circular and elliptical irises computed using the two methods described in this paper as well as the theoretical methods of Eq. \ref{eq:pol_love} and Eq. \ref{eq:waveguide_rad_reaction}. As shown, excellent agreement between the analytical expressions and the numerical extractions is obtained, verifying the proposed methods. Since the circular iris considered here does not possess a resonance, it is expected that the dynamic polarizabilities extracted numerically  are well-approximated by the theoretical expressions. Next, we examine the case of an elliptical iris, as shown in Fig.\ref{fig:pol_results_rwg}b. The extracted polarizabilities computed using the two numerical extraction methods of previous section exhibit excellent agreement. However, the analytical exhibits significant deviation from the numerical methods. This is expected since the elliptical iris supports a resonance over the frequency band of interest, which is not captured in the analytical expressions derived for the static field. This case further highlights the need for a precise numerical method to compute the polarizability of a metamaterial element.

It is worth noting that an elliptical iris etched in a rectangular waveguide is well-known in the antenna engineering community as the unit cell of a slotted waveguide antenna (SWA). SWAs are particularly attractive due to their advantages in terms of design simplicity, weight, volume, power handling, directivity, and efficiency \cite{oliner1993leaky,grbic2002leakywave,gomez2005design,maci2005pole,sievenpiper2005forward,grbic2002leakywave}. SWAs rely on the gradual leakage of the guided mode through the slots. The metasurfaces considered throughout this paper also share this feature, since they also leak energy from the guided wave through the metamaterial elements. In other words, the methodology developed in this paper can also be applied to leaky wave antennas for modeling, simulation, and designing \cite{pulido2016discrete}.

\begin{figure}
\centering{
\includegraphics[width=\columnwidth]{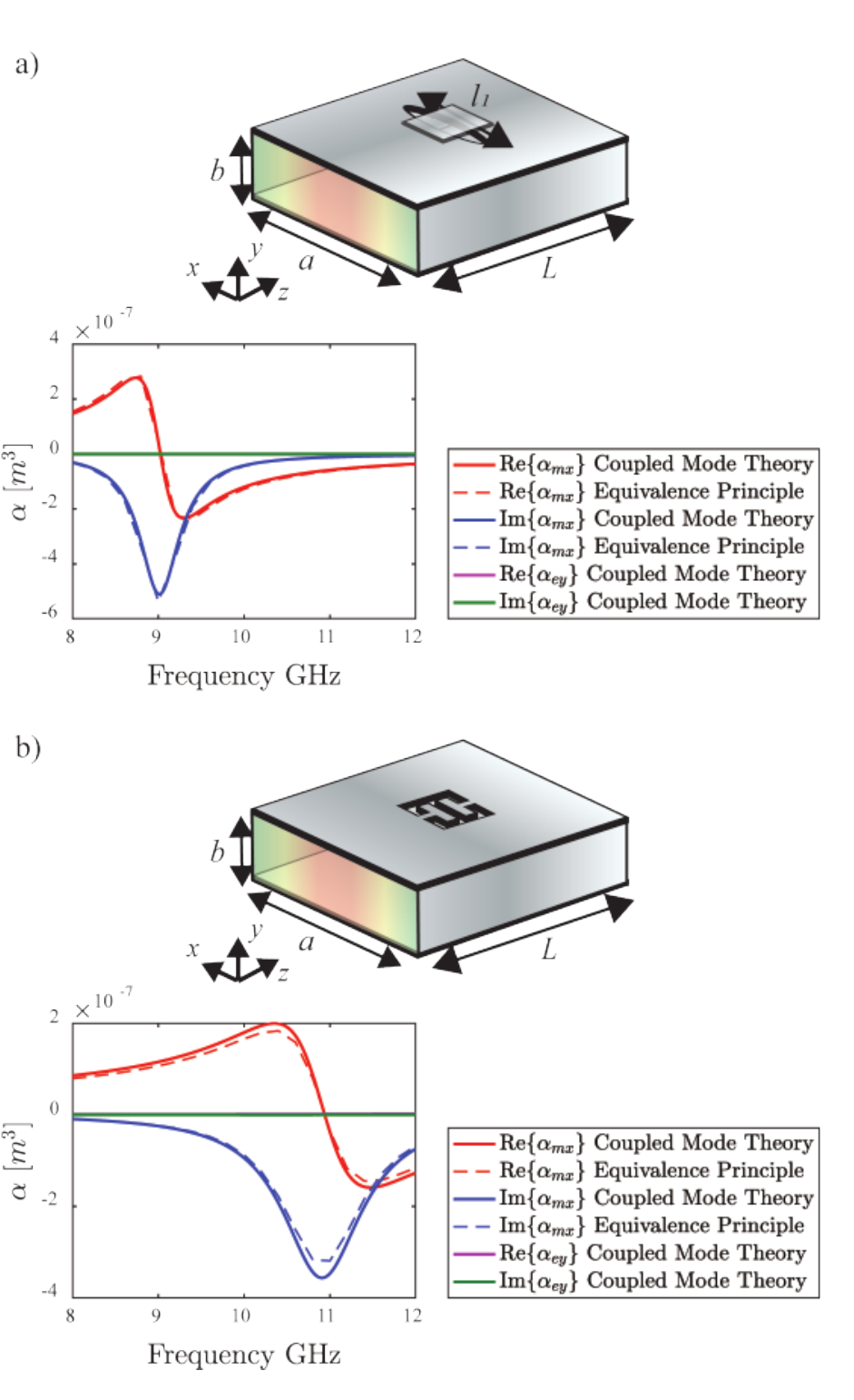}}
\caption{Dynamic Polarizability of small apertures. a) Iris-fed patch. For this example, The iris length is $l_1=6$ mm and the square patch size is $l_s=5$ mm. b) cELC resonator. Its dimensions are $l_1=5$ mm and $l_2=5$ mm, and a thickness of $c=1$ mm.}\label{fig:pol_results_rwg2}
\end{figure}

While the circular and the elliptical irises may be analyzed using analytical expressions for the polarizabilities derived in the static limit, such closed-form expressions are not available for most metamaterial designs. For example, an element of potential interest in the design of metasurface antennas is the iris-fed patch, shown in Fig.\ref{fig:pol_results_rwg2}a \cite{stevenson2016mtenna}. The inclusion of the metallic patch above the iris enhances the resonant response of the element, as exemplified by the narrower and stronger resonant response. Another common metamaterial element is the cELC, shown in Fig.\ref{fig:pol_results_rwg2}b, commonly used in metasurface antenna designs. The resonant response of the cELC is highly susceptible to variations in its geometry \cite{sleasman2015element,odabasi2013electrically,yoo2016efficient}. For both elements, we observe excellent agreement between the two numerical methods of equation Eq. \ref{eq:pol_love} and equation Eq. \ref{eq:pol_rwg}, as shown in Fig. \ref{fig:pol_results_rwg} and \ref{fig:pol_results_rwg2}.

The geometry of metamaterial elements can be quite complicated, such that the numerical integrals required in Eq. \ref{eq:pol_love} are likely to yield inaccuracies. For this reason, the extraction based on computing the waveguide scattering parameters is likely to be more reliable and easier to implement; moreover, this method can also be used in measurements on fabricated samples. It worth noting that in all of the results presented in Fig.\ref{fig:pol_results_rwg} and \ref{fig:pol_results_rwg2}, it can be seen the electric polarizability is much smaller than the magnetic polarizability---in fact, three orders of magnitude smaller. This phenomenon is expected considering that the geometry under study corresponds to a small opening in a metallic wall. 


\section{\label{sec:Planar-Waveguide} Polarizability Extraction in a Parallel Plate Waveguide}

In this section, we consider the case of a 2D waveguide and examine a metamaterial element that is etched on one wall of a parallel plate waveguide. While the polarizability extraction method based on the equivalence principle holds for the 2D waveguide-fed element, the nature of the waveguide modes changes substantially from the formulation in section \ref{sec:PolarizabilityExtraction}. In this section, we modify this extraction technique to be applicable to planar waveguide systems. As previously described in section \ref{sec:PolarizabilityExtraction}, a metamaterial element scattering into a waveguide can be described in terms of a sum of waveguide modes. Because the element is placed in the upper surface of the waveguide, the boundary condition dictates the tangential electric field and the normal magnetic field to be zero and the element can only couple to the transverse magnetic (TM) modes. Since the natural symmetry of the system is cylindrical, mode decomposition is simpler if we use cylindrical coordinates ($r,\theta$). 

Setting the origin of the coordinate system to the center of the metamaterial element, the \(z\)-components of the scattered electric field---for the TM modes characterized by the $(m,n)$ indices--- are given by
\begin{subequations}
\begin{align}
&E^{sc}_{z,c}=\frac{\beta_m}{k} H_n^{(2)}(\beta_m r)\cos(n\theta) \\
&E^{sc}_{z,s}=\frac{\beta_m}{k} H_n^{(2)}(\beta_m r)\sin(n\theta)
\end{align}
\end{subequations}
\noindent where the subscripts ``c" and ``s" refer to modes that have angular dependence \(\cos(n\theta)\) and \(\sin(n\theta)\), respectively. The propagation constant is given by \(\beta_m=\sqrt{k^2-(m\pi/h)^2}\), where \(h\) is the height of the waveguide. Invoking the superposition principle, the total solution for the \(z\)-component of the scattered electric field can be expressed as

\begin{equation}\label{eq:Ezmn_pwg}
E_z=\sum_{n}\sum_{m}A_{mn}^s E_{z,s}^{mn}+A_{mn}^c E_{z,c}^{mn}
\end{equation}
When \(h<\pi/k\), only the \(m=0\) mode is propagating, and in this case the electric field at all points where \(r\gg h/\pi\) is dominated by the \(m=0\) mode. Therefore we can reduce Eq. \ref{eq:Ezmn_pwg} to 
\begin{equation}
E_z=\sum_{n}A_{n}^s E_{z,s}^{0n}+A_{n}^c E_{z,c}^{0n}.
\label{eq:Ez-field}
\end{equation}
The \(m=0\) modes are given by 
\begin{subequations} \label{eq:ppwg-eigenmodes}
\begin{align}
&E_{z,c}^{0n}=H_n^{(2)}(kr)\cos(n\theta) \\
&E_{z,s}^{0n}=H_n^{(2)}(kr)\sin(n\theta).
\end{align}
\end{subequations}

\begin{figure}
\centering{
\includegraphics[width=\columnwidth]{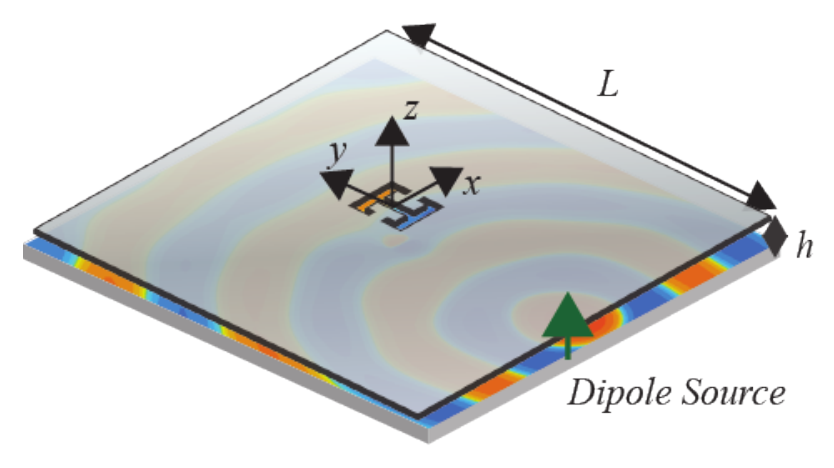}}
\caption{Planar waveguide with a CELC etched at the center. The traveling wave is excited by an electric dipole oriented along the $z-$direction. The waveguide dimensions are $L=100 mm$, $h=1.27 mm$. Dipole Source Location $0.45L$ \label{fig:panel-single}}
\end{figure}

The amplitude coefficients, $A_n$, can be found from the scattered electric field \(E_z\) using the orthogonality of the \(\{\sin(\theta),\cos(\theta)\}\) basis. By integrating over a circle of radius \(r\) centered at the origin of the metamaterial element, as shown in Fig.\ref{fig:panel-single}, it is possible to define the amplitude coefficients as

\begin{subequations}\label{eq:Amp-Esc-pwg}
\begin{align}
&A_n^s=\lim_{r \rightarrow \infty} \frac{1}{\pi H_n^{2}(kr)}\int_0^{2\pi} E_z(r,\theta)\sin(n\theta) d\theta \\
&A_n^c=\lim_{r \rightarrow \infty} \frac{1}{\pi(1+\delta_{n0})H_n^{2}(kr)}\int_0^{2\pi} E_z(r,\theta)\cos(n\theta) d\theta.
\end{align}
\end{subequations}

To better illustrate the utility of equations Eq. \ref{eq:Amp-Esc-pwg}, we consider a lossless parallel plate waveguide, fed by a cylindrical source oriented in the $z-$direction,, as shown in Fig. \ref{fig:panel-single}. The source is placed far enough from the metamaterial element to avoid evanescent coupling. A single cELC resonator is assumed to be patterned on the center of the top plate of the waveguide, with the same geometrical parameters as the cELC presented in the previous section. Since the full-wave simulation domain represents the total field instead of the scattered field,  this structure is simulated with and without the cELC, and the difference of the two simulation results are taken to obtain the scattered field due to the metamaterial element, such that $E_z^{sc}= E_z^{tot} -E_z^{0}$ at the plane $z=h/2$. Once the scattered field is computed, the integration outlined in Eq. \ref{eq:Amp-Esc-pwg} is performed to find the amplitude coefficients. The integration radius is selected electrically large enough so that the evanescent modes have decayed ---it is also ensured the integration curve does not contain the cylindrical source. 
\begin{figure}
\centering{
\includegraphics[width=0.9\columnwidth]{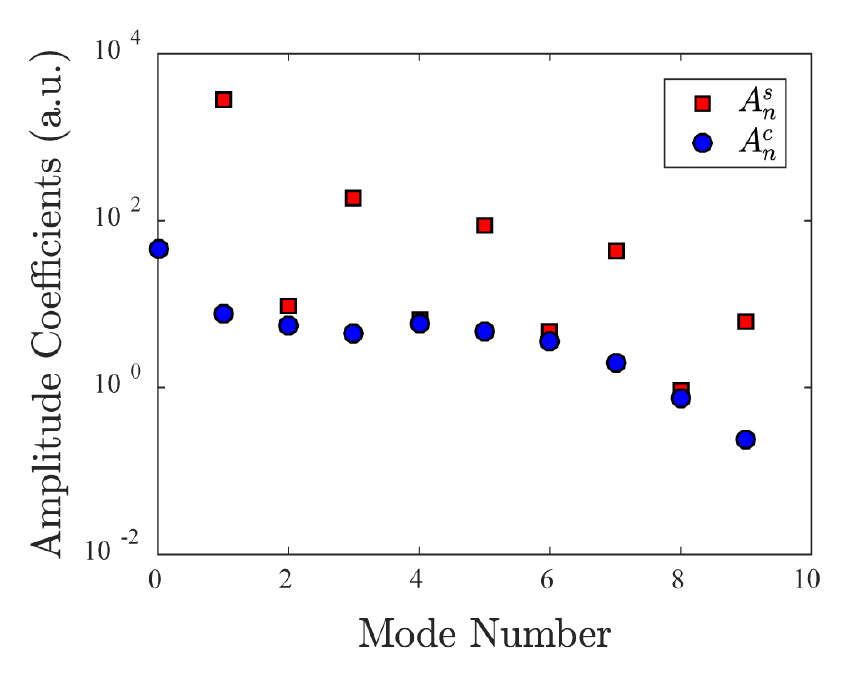}}
\caption{Amplitude Coefficients of the scattered field $E_z$. The scattered fields from the metamaterial element are added to the incident field produced by an electric current source. \label{fig:amp-coeff}}
\end{figure}

Figure \ref{fig:amp-coeff} shows the magnitude of the amplitude coefficients for the scattered fields computed for the meta-atom shown in Fig.\ref{fig:panel-single}. To better illustrate the physics behind these coefficients, we apply Poynting's theorem Eq. \ref{eq:poynting} which directly links the amplitude coefficients and the dominant dipole moments of the metamaterial element. In contrast to the 1D case examined in section \ref{sec:PolarizabilityExtraction}, the amplitude coefficients are not related to the scattering parameters, but rather to the scattered fields, by means of Eq. \ref{eq:Amp-Esc-pwg}. Moreover, while the location of the metamaterial element in the rectangular waveguide limits the calculation of a single component of the magnetic polarizability, such limitation disappears in the case of the planar waveguide. 

The cylindrical wave propagating through the waveguide may excite the two tangential components of the magnetic polarizability, which leads to a more complete characterization of the polarizability tensor of the element. The scattered fields generated by the metamaterial element can be represented as the sum of the moments of the surface current $\boldmath{J}^n_m$ multiplied by the different eigenmodes of the scattered fields shown in Eq. \ref{eq:ppwg-eigenmodes}, more explicitly, this relationship is given by 
\begin{equation}\label{eq:Ez_Dipoles}
E_z=\frac{m_xZ_0k^2}{4h} E_{z,s}^{01}+\frac{m_yZ_0k^2}{4h} E_{z,c}^{01}+\frac{-ip_zk^2}{4h\epsilon_0} E_{z,c}^{00}
\end{equation}
A direct mapping between Eq.\ref{eq:Ez_Dipoles} and Eq. \ref{eq:Amp-Esc-pwg} demonstrates that the first three amplitude coefficients, \(\{A_0^c, A_1^c, A_2^s\}\) are directly related to the three dominant dipole moments as \cite{bowen2017effective}

\begin{equation}\label{eq:peff_meff_pwg}
m_x=A_1^s\frac{4h}{Z_0 k^2} \quad m_y=A_1^c\frac{4h}{Z_0 k^2}\quad p_z=A_0^c\frac{i4h\epsilon_0}{k^2}.
\end{equation}

As shown in Fig. \ref{fig:amp-coeff}, the predominant amplitude mode is $A^s_1$, which is directly associated with $m_x$, while the amplitude of the modes $A^c_0$ and $A^c_1$, associated with $p_z$ and $m_y$, are significantly smaller---by two orders of magnitude. The effective polarizabilities given the incident wave due to the line source, can be directly obtained from their corresponding dipole moments given in Eq. \ref{eq:peff_meff_pwg} as
\begin{equation}
\label{eq:alpha_pwg}
\alpha_{z}^p=p_z/E_z^{0}\quad
\alpha_{xy}^m=m_y/H_x^{0} \quad
\alpha_{xx}^m=m_x/H_x^{0}.
\end{equation}

\begin{figure}
\centering{
\includegraphics[width=\columnwidth]{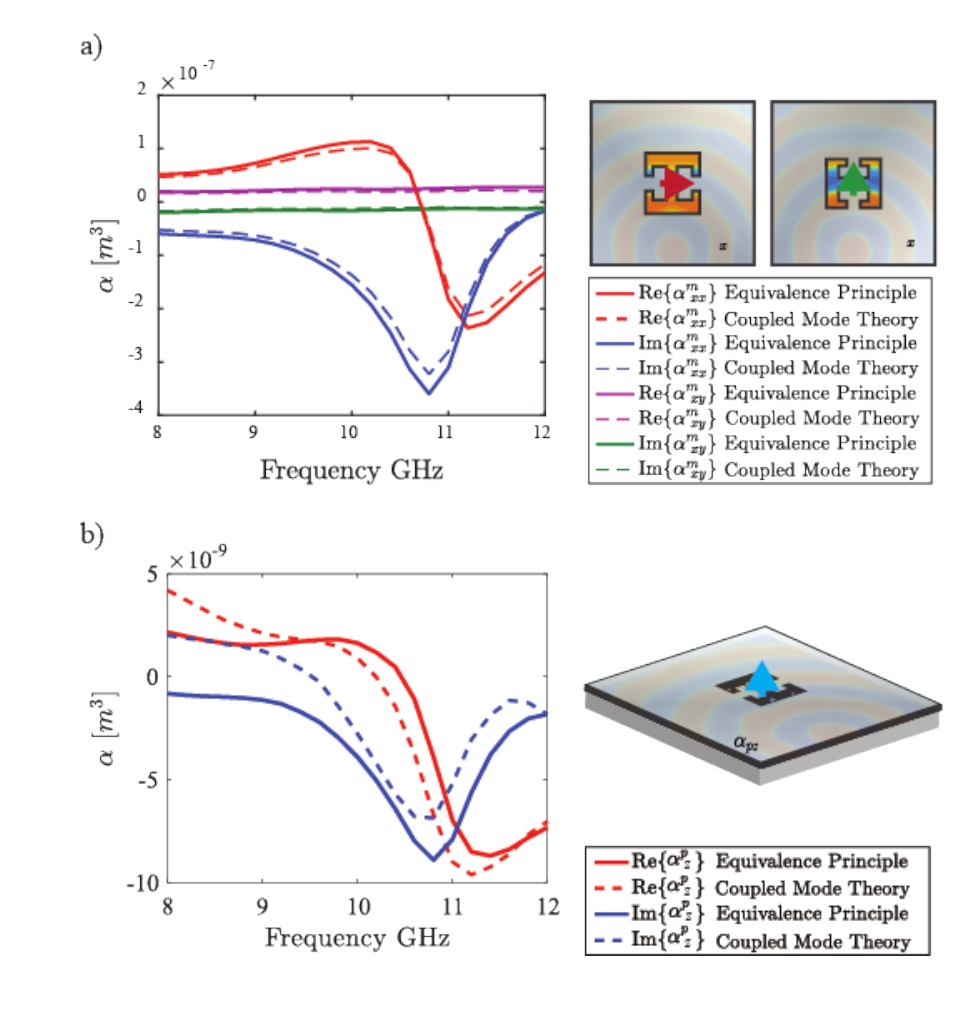}}
\caption{Effective magnetic polarizabilities calculated for the ELC embedded in the planar waveguide. The metamaterial element size is enlarged in the figure to clarify its orientation with respect to the incident field.}\label{fig:pols-PWG}
\end{figure}

For clarification, the double indices on the polarizabilities represent the entry in the polarizability tensor; for example, $\alpha_{xy}$ represents the component of the polarizability that generates a dipole moment oriented in $y$, due to the $x$ component of the incident field. In order to find all three components of this tensor, it is necessary to rotate the cELC by $\pi/2$, and perform the same extraction technique. The electric polarizability and two of the components of the magnetic polarizability tensor are thereby obtained and shown in Fig.\ref{fig:pols-PWG}a. As shown, excellent agreement for the magnetic polarizabilities is obtained between the two numerical polarizability extraction methods. For this particular example, note that the polarizability $\alpha^m_{yy}$ has a resonant response out of the X-band, but its geometry can be modified such that it has both resonances in the same band \cite{yoo2016efficient}. In the case of the electric polarizability (Fig.\ref{fig:pols-PWG}b), the numerical values obtained are significantly smaller, which makes it susceptible to numerical inaccuracies when the integration in Eq. \ref{eq:Amp-Esc-pwg} is performed. 

\begin{figure*}
\centering{
\includegraphics[width=0.85\textwidth]{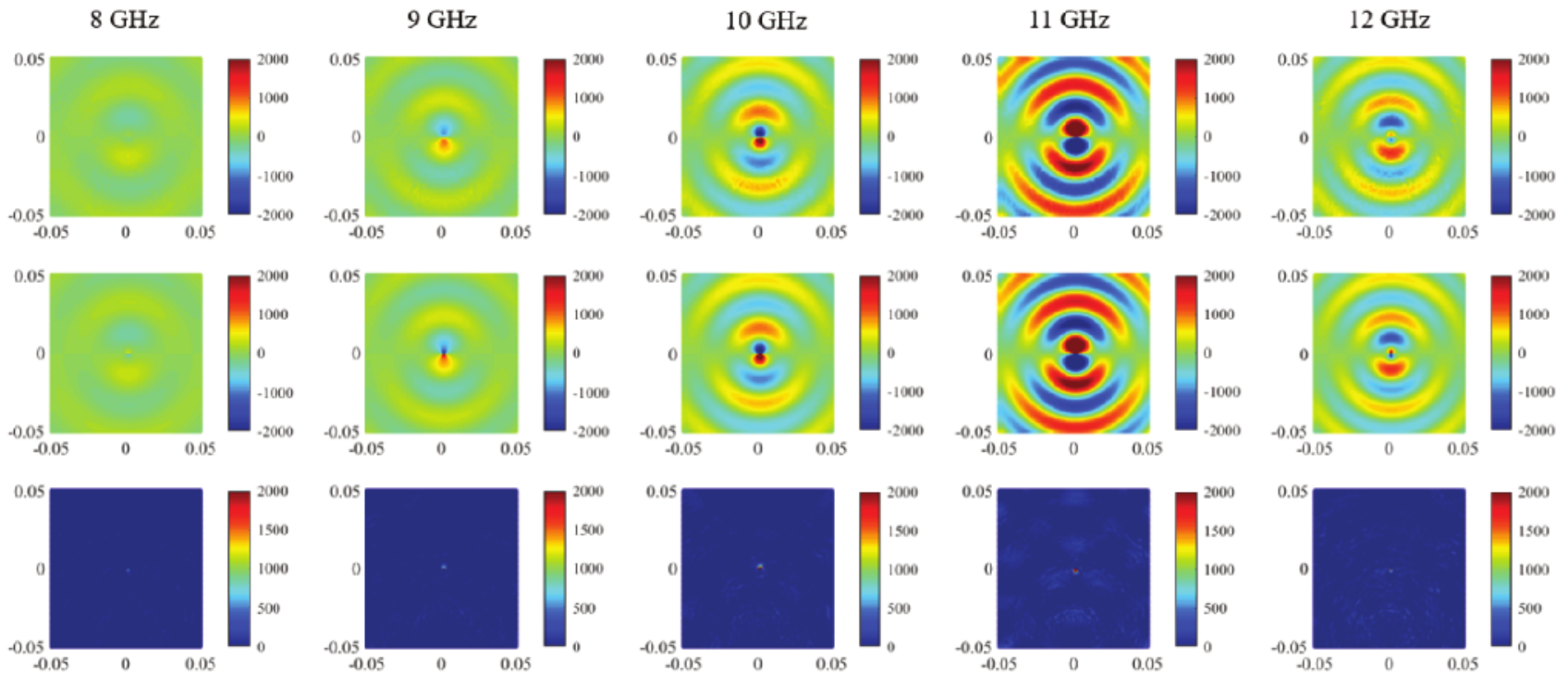}}
\caption{Real part of the scattered field $E_z (V/m)$ at different frequencies. (Top row) Full-wave simulation in \textit{CST Microwave Studio}. (Middle Row) Analytic expression from Eq. \ref{eq:Ez-field} up to the dipolar term only. The difference is shown in the bottom row.}\label{fig:scattered-fields}
\end{figure*}

It is important to highlight in this example that the term $A^s_3$ is associated with the quadrupole moment. This term has been traditionally neglected in most metamaterial design strategy. The example at hand provides a useful framework to examine the contribution of the quadrupole term and the error introduced by neglecting it. To study the impact of the quadrupole term, we compared the simulated scattered field (shown in Fig. \ref{fig:scattered-fields} first row) with its theoretical expression Eq. \ref{eq:Ez-field} up to only the dipolar contribution, as shown in the second row of Fig.\ref{fig:scattered-fields}. We observe excellent agreement between the two rows, confirming the assumption that the main contribution of the scattered field is dipolar. The physical implications of this result can be understood by calculating the difference between the scattered fields from a full-wave simulation and from the analytical expression in (Eq. \ref{eq:Ez-field}). As shown in the third row of Fig. \ref{fig:scattered-fields}, the error due to assuming the dominant dipolar term is several orders of magnitude smaller than the amplitude of the scattered field, and the largest discrepancy is observed within the close vicinity of the metamaterial element. This result, in conjunction with the amplitude coefficients shown in Fig.\ref{fig:amp-coeff} also demonstrates that most of the radiation is associated with the dipolar term and higher order modes can be ignored. However, if the elements are placed at distances where these higher order modes have not decayed, they can alter the coupling between the two meta atoms and change the total scattered fields inside the waveguide. 

Another interesting point to highlight is that while the geometric characteristics of the cELC used in Fig.\ref{fig:pol_results_rwg2}c and \ref{fig:panel-single} are the same, its resonant response as manifested by the effective polarizability changes depending on the host waveguide, as can be observed by comparing Fig.\ref{fig:pol_results_rwg}d and Fig.\ref{fig:pols-PWG}a. Therefore, the electromagnetic response of a metamaterial element embedded in a waveguide not only depensd on its intrinsic geometrical characteristics, but also depends on the waveguide geometry where it is inserted. 

\section{\label{sec: Conclusions} Conclusions}

In this paper we have presented a comprehensive method for extracting the effective polarizabilities of metamaterial elements patterned in 1D rectangular and 2D parallel plate waveguides. In our approach, the S-parameters---along with the knowledge of the normalized fields inside the waveguide---are used to find the electric and magnetic polarizabilities, making this technique a powerful tool for the design and characterization of metasurfaces. To demonstrate the validity of our approach, the polarizability extraction method was compared with direct extraction of the tangential components of the scattered fields, and excellent agreement between the two methods was demonstrated. We have also shown that the dipole modes predominate the scattering from metamaterial elements, since all higher-order multipole fields decay more rapidly with distance. The concepts presented in this paper pave the way for a simple and efficient approach to the analysis of metasurface antennas. The combination of polarizability extraction techniques with the dipole model provides a powerful and inherently multiscale modeling tool to design and characterize metasurface structures without any limitation on the element geometry or periodicity assumptions common to other homogenization techniques. 

\appendix
\section{Radiation Reaction inside a Rectangular Waveguide \label{ap:rad-reaction}}
 As described in section \ref{sec:DDA}, the radiation reaction in a rectangular waveguide can be found by taking the real part of the surface integral of the Pointing's vector $\mathbf{S}$. Let us consider again a thought experiment where two collocated electric and magnetic dipole $\mathbf{p}$ and $\mathbf{m}$ are placed at position $\mathbf{r}_0$,in an environment that is described by the Green's function $\mathbf{G}(\mathbf{r}; \mathbf{r}_0)$. The surface integral of the Poynting's vector is given by
\begin{equation}\label{eq:ap-poynting}
\int \mathbf{S}\cdot d\mathbf{a} = i\omega (\mathbf{p^*}\cdot\mathbf{E} -  \mu_0 \mathbf{m^*}\cdot \mathbf{H})\\
\end{equation}
 \noindent where $\mathbf{E}$ and $\mathbf{H}$ correspond to the total fields inside the waveguide. Evaluating such fields at the dipole's location we get
\begin{subequations}
 \begin{align}
\int \mathbf{S}\cdot d\mathbf{a}=i\omega (\mathbf{p^*}\cdot \mathbf{E} -  \mu_0 \mathbf{m^*}\cdot \mathbf{H}) \nonumber \\
 = i\omega (\mathbf{p*}\cdot \mathbf{G}_{ee}(r_0,r_0)\cdot\mathbf{p} -  \mu_0 \mathbf{m^*}\cdot\mathbf{G}_{mm}(r_0;r_0)\cdot\mathbf{m})
 \end{align}\label{eq:integral_green0}
\end{subequations}
Now, considering the real part of Eq. \ref{eq:integral_green0} it is possible to obtain a direct relationship between the total power radiated and the imaginary components of the Green's functions as
\begin{subequations}
 \begin{align}
\mathrm{Re}\left\lbrace \int \mathbf{S}\cdot d\mathbf{a} \right\rbrace = \omega |\mathbf{p}|^2 \mathrm{Im}\left\lbrace G_{ee}(r_0,r_0) \right\rbrace \nonumber \\ 
- \mu_0 \omega |\mathbf{m}|^2 \mathrm{Im}\left\lbrace G_{mm}(r_0,r_0) \right\rbrace
 \end{align}\label{eq:integral_green0-real}
\end{subequations}
On the other hand, from the modal expansion of the fields Eq. \ref{eq:Ep} and Eq. \ref{eq:Em}, the same integral shown in Eq. \ref{eq:ap-poynting} results in
\begin{equation}\label{eq:integral_modes}
  \int \mathbf{S}\cdot d\mathbf{a} =\frac{1}{Z_n}(|A_n^+|^2 + |A_n^-|^2).
\end{equation}
\noindent where $Z_n$, $A_n^+$ and $A_n^-$ have been previously defined in \ref{sec:PolarizabilityExtraction}. By using the equations for the amplitude coefficients shown in Eq. \ref{eq:Ap-Am-2} into (Eq. \ref{eq:integral_modes}) we obtain
\begin{subequations}
\begin{align}
(|A_n^+|^2 + |A_n^-|^2) = \frac{\omega^2 Z_n^2}{4} (|E_n^+|^2 |p|^2 - \mu_0|H_n^+|^2 |m|^2 ) 
\end{align}
\end{subequations}
In addition, it was previously demonstrated that for the fundamental mode, the fields $E_n^+$ and $H_n^+$ are normalized by means of Eq. \ref{eq:ALLfields}. Replacing the specific expressions for the normalized fields, and equating Eq. \ref{eq:integral_green0} and Eq. \ref{eq:integral_modes} we obtain a direct relationship between the amplitude of the normalized modes and the imaginary Green's functions, which are our ultimate goal in this derivation. More explicitly,
\begin{subequations}
\begin{align}
\omega |\mathbf{p}|^2 \mathrm{Im}\left\lbrace G_{ee}(r_0,r_0) \right\rbrace  
- \mu_0 \omega |\mathbf{m}|^2 \mathrm{Im}\left\lbrace G_{mm}(r_0,r_0) \right\rbrace \nonumber\\
= \frac{\omega^2 Z_n^2}{4} (|E_n^+|^2 |p|^2 - \mu_0|H_n^+|^2 |m|^2 ) 
\end{align}\label{eq:rad_reac_identity}
\end{subequations}
It can be observed that, in order to satisfy Eq. \ref{eq:rad_reac_identity} the terms multiplying the magnitude of the dipole moments $|p|^2$ and $|m|^2$ must be equal. After some algebraic derivation it is possible to conclude that
\begin{equation}\label{eq:greens0}
\mathrm{Im}\left\lbrace G_{ee}(0) \right\rbrace = \frac{k^2}{\beta ab}  \quad \mathrm{Im}\left\lbrace G_{mm}(0) \right\rbrace= \frac{k}{ab}.
\end{equation} 
\bibliography{polarizability_references}
\end{document}